# Towards Highly Efficient State Estimation with Nonlinear Measurements in Distribution Systems

Ying Zhang, *Student Member, IEEE*, Jianhui Wang, *Senior Member, IEEE*

*Abstract*—This letter proposes a novel and highly efficient distribution system state estimation (DSSE) algorithm with nonlinear measurements from supervisory control and data acquisition (SCADA) systems. Conventional DSSE, *i.e.*, a weighted least square (WLS)-based method, requires multiple Monte Carlo simulations for overall accuracy evaluation and has high calculation cost due to the nonlinear iterative process. The proposed method uses the Taylor series of voltages for constructing a linear DSSE model in the interval form and then solve this model by interval arithmetic. Compared with the nonlinear WLS-based method, the proposed method obtains accurate and robust estimates via a single random sampling of measurements and is computationally efficient. The comparative analysis of the IEEE 34-bus distribution system points to improved estimation results relative to the WLS-based method.

## I. Introduction

DISTRIBUTION systems are undergoing radical changes in operation and control due to renewable integration, which emphasizes the importance of distribution system state estimation (DSSE) [1]. The DSSE procedure converts redundant meter readings and other available information into an estimate of system states. The measurements in DSSE can be the voltage magnitudes, power injections, and power flows from supervisory control and data acquisition (SCADA) systems, or voltage and current phasor recorded by phasor measurement units (PMUs) [2]. State estimation widely uses Monte Carlo simulations (MCSs) [3] to obtain random samplings of measurements including noises owing to the assumption that these noises follow Gaussian distributions. Also, traditional nonlinear DSSE methods adopt the Gauss-Newton method based on the weighted least square (WLS) criterion to perform the iterative estimation process. On the other hand, to correctly evaluate the estimation performance, the number of the required samples in MCSs is tremendous, which brings a heavy computation load to the WLS-based methods [1].

To mitigate the deficiency of MCSs, analytical methods such as nonlinear programming [4], [5] and interval arithmetic [6] are proposed to provide the upper and lower bounds of all possible state variables that meet all constraints from measurements. For instance, the authors of [4] used a constrained nonlinear programming approach to obtain the ranges of states in transmission systems, while [5] extends this

Y. Zhang and J. Wang are with the Department of Electrical and Computer Engineering, Southern Methodist University, Dallas, TX, USA 75205 (e-mail: yzhang1@smu.edu; jianhui@smu.edu).

boundary optimization method to distribution systems with PMU installation. For higher computational efficiency, [6] solves a linear DSSE model by interval arithmetic and obtains accurate state estimates. However, the methods in [5] and [6] require the installation of PMUs in distribution systems. Due to lack of PMUs in some practical distribution systems, the metering data recorded by SCADA systems and pseudo-measurements collected at loads or distributed generators (DGs) are widely used in the existing DSSE methods [2]. Moreover, these measurements lead to nonlinear DSSE models that are iteratively solved, and thus the process is time-consuming. This letter presents a highly efficient DSSE method with these nonlinear measurements to handle the uncertainty of random measurement noises. The main contributions of the proposed method are concluded as (i) constructing a novel DSSE model to avoid multiple runs of the conventional WLS-based DSSE procedure and (ii) accelerating accurate state estimates by interval arithmetic without the use of PMU data.

## II. Methodology

### A. WLS-based Method

In classical state estimation, the relationship between measurements and state variables are depicted as
$$z = h(x) + e \tag{1}$$
where the state vector $x \in \mathbb{R}^{n \times 1}$, and the measurement vector $z \in \mathbb{R}^{m \times 1}$, $m \geq n$; $h(x)$ is a measurement function about $x$; $e$ denotes the vector of measurement noises that usually obey Gaussian distributions, *i.e.*, $e \sim N(\mathbf{0}, \mathbf{R})$, and $\mathbf{R}$ denotes the covariance matrix of these measurement noises.

The WLS criterion is used to minimize the sum of weighted measurement residuals, $J$:
$$J = [z - h(x)]^T W [z - h(x)] \tag{2}$$
where $W$ represents the weight matrix of these measurements, and $W = R^{-1}$.

Optimal states are iteratively solved by the Gauss-Newton method until $\Delta x$ in adjacent iterations is sufficiently small.
$$\partial J / \partial x = H(x)^T W [z - h(x)] = \mathbf{0} \tag{3}$$
$$\Delta x = (H(x)^T W H(x))^{-1} H(x)^T W [z - h(x)] \tag{4}$$
$$x^{(t+1)} = x^{(t)} + \Delta x \tag{5}$$
where $H(x)$ is the Jacobian matrix and $H(x) = \partial h(x)/\partial x$.

In the conventional DSSE, SCADA systems provide the metering data of voltage magnitudes and powers, and pseudo-measurements are also used to achieve the system observability. Also, the substation acts as a phase reference, and the state variables are the voltage magnitudes and phase angles at all buses. The formulation of the nonlinear measurement functions $h(x)$ can be found in [2] and references therein.

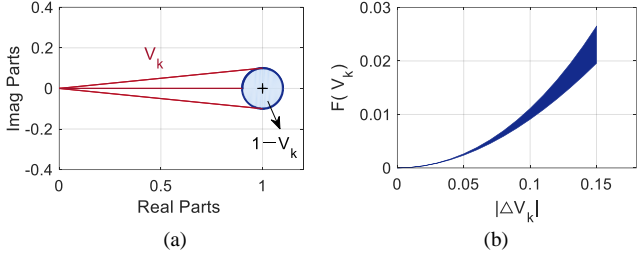

Fig. 1. Schematic diagram of the proposed linearization. (a) The complex plane of $V_k$. (b) Approximation loss $\mathrm{F}(V_k)$

## B. Proposed DSSE Algorithm

This section proposes a linear approximation for efficient DSSE formulation and the solving method of the linear model via interval arithmetic.

Define $\Delta V_k = 1 - V_k$, where $V_k$ denotes the voltage phasor at bus $k$. Considering the small voltage drops along the distribution lines and the normal voltage limits (0.95~1.05 p.u.) in practical systems [6], apply the Taylor series of $\Delta V_k$ around zero as $1/(1 - \Delta V_k) = \sum_{n=0}^{+\infty}(\Delta V_k)^n$. Further, the following equation is obtained by ignoring the high order terms [7]:

$$1/(1 - \Delta V_k) \approx 1 + \Delta V_k = 2 - V_k \quad (6)$$

Fig.1 depicts the accuracy loss introduced by (1) as $\mathrm{F}\ V_k = |1/V_k - 2 - V_k|$, and for instance, the error for $|\Delta V_k| = 0.1$ is around 0.01. The approximation relationship $1/V_k = 2 - V_k$ is used for constructing the measurement functions in the proposed DSSE model, which are shown below.

1) The power flow measurements $P_{ik}$ and $Q_{ik}$ at branch $i-k$ are expressed as

$$S_{ik} = P_{ik} + jQ_{ik} = V_i[y_{ik}(V_i - V_k)]^* \quad (7)$$

where $P_{ik}$ and $Q_{ik}$ denote the real and reactive powers at this branch; $y_{ik}$ denotes the nodal admittance between buses $i$ and $k$, and the function $[\cdot]^*$ represents the complex conjugate; $V_i$ and $V_k$ denote the voltage phasors at these two buses.

Apply (6) to (7), and a closed-form expression is obtained.

$$S_{ik}(2 - V_i) = [y_{ik}(V_i - V_k)]^* \quad (8)$$

Formula (8) can be linearly expressed as $S_{ik}V_i + y_{ik}{}^*V_i{}^* - y_{ik}{}^*V_k{}^* = 2S_{ik}$.

2) The measurement function of the power injection at bus $k$ holds below, and similar to (8), further expressed as a linear one.

$$S_k = P_k + jQ_k = V_k \sum_{l \in \mathcal{N}(k)} [y_{lk}(V_l - V_k)]^* \quad (9)$$

$$S_k(2 - V_k) = \sum_{l \in \mathcal{N}(k)} [y_{lk}(V_l - V_k)]^* \quad (10)$$

where $P_k$ and $Q_k$ denote the real and reactive powers at bus $k$, and $\mathcal{N}(k)$ is the set of all buses connected to bus $k$.

3) The measurement function of voltage magnitudes at bus $k$ is approximated as

$$|V_k| = \sqrt{V_{k,r}^2 + V_{k,x}^2} \approx V_{k,r} \quad (11)$$

where $V_{k,r}$ and $V_{k,x}$ represent the real and imaginary parts of voltages at bus $k$, and the small angle differences of distribution lines are considered, e.g., 0.1 degrees per mile [6].

Reorganize the linear equations (8), (10) and (11) as

$$BV + DV^* = E \quad (12)$$

where $V$ denotes the vector of the nodal voltage phasors, and $B$, $D$, and $E$ are the corresponding coefficient matrices and suppressed here due to the limited space. Further, express (12) in rectangular coordinates as

$$\begin{bmatrix} B_r + D_r & -B_x + D_x \\ B_x + D_x & B_r - D_r \end{bmatrix} \begin{bmatrix} V_r \\ V_x \end{bmatrix} = \begin{bmatrix} E_r \\ E_x \end{bmatrix} \quad (13)$$

where the subscripts $r$ and $x$ denote the real and imaginary parts of complex numbers. For simplicity, (13) is expressed as

$$Ax = b \quad (14)$$

where $x = \begin{bmatrix} V_r \\ V_x \end{bmatrix}$ denotes the state vector; $A = \begin{bmatrix} B_r + D_r & -B_x + D_x \\ B_x + D_x & B_r - D_r \end{bmatrix}$, and $A \in \mathbb{R}^{m \times n}$; rank $A = n$, i.e., a full rank; $b = \begin{bmatrix} E_r \\ E_x \end{bmatrix}$ and $b \in \mathbb{R}^{m \times 1}$.

Note that (14) does not consider measurement noises and involves various levels of approximation on the voltage magnitudes and powers. Next, based on (14), we use interval arithmetic [8] to handle the accuracy loss introduced by the linear approximation and these measurement noises in DSSE. We consider the measurement noises by updating (14) to an interval equation, where an interval number is defined as $[a] = [a_l, a_u] = \{a \in \mathbb{R} \mid a_l \leq a \leq a_u\}$, and interval vectors and matrices are constructed similarly. According to the $3\sigma$ rule of a Gaussian distribution, where $\sigma$ denotes the standard deviation, 99.73% of values from the distribution are within three times of standard deviations [2]. Hence, the maximum measurement errors (i.e., $\mp 3\sigma$) are superposed onto the corresponding measurements to obtain the lower and upper bounds of $A$ and $b$, i.e., $[A]$ and $[b]$. By this relaxation, these measurement intervals enclose their true values.

For the interval equation $[A][x] = [b]$, the interval solution hull $[x]$ is the interval vector with the smallest radius containing all possible solutions of $Ax = b$, where $A_l \leq A \leq A_u$ and $b_l \leq b \leq b_u$. Also, an equivalence relation with a dummy vector $[y]$ exists below:

$$[A][x] = [b] \iff \begin{bmatrix} [A] & -I \\ 0 & [A]^T \end{bmatrix} \begin{bmatrix} [x] \\ [y] \end{bmatrix} = \begin{bmatrix} [b] \\ 0 \end{bmatrix} \quad (15)$$

where $I \in \mathbb{R}^{m \times m}$ is an identity matrix, and $[y] \in \mathbb{R}^{m \times 1}$. Compactly express (15) as $[\mathcal{A}][X] = [\mathcal{B}]$, and the interval symbol [ ] is omitted below, e.g., the augmented state vector $X = [x\ y]^T$.

Equation (15) is solved by a Krawczyk-operator algorithm [6]. An initial interval solution $X^{(0)}$ is calculated by

$$X^{(0)} = ([-\alpha, \alpha], ..., [-\alpha, \alpha])^T \quad (16)$$

where $\alpha = \frac{\|C\mathcal{B}\|_\infty}{1 - \beta}$ and $\beta = \|I - C\mathcal{A}\|_\infty$; $C$ is a preconditioning point matrix, $C^{-1} = Mid[\mathcal{A}]$, and $Mid[\cdot]$ denotes the midpoint of an interval; $\|\cdot\|_\infty$ denotes the infinite norm of a vector, and this initial solution $X^{(0)}$ contains the final solution.

The following process at iteration $j$ is used to gradually approach the final solution hull until $\|X^{(j+1)} - X^{(j)}\|_\infty \leq \varepsilon$:

TABLE I
MEASUREMENT ARRANGEMENT IN TEST SYSTEM

| Measurements | | Location |
|---|---|---|
| SCADA | $|V|$ | 800, 820, 832, 834 |
| | $P, Q$ | 800-802, 808-812, 824-828, 832-858, 860-836 |
| Pseudo-meas. $P, Q$ | | All load nodes and DG nodes |

TABLE II
COMPARISON IN ESTIMATION ACCURACY AND COMPUTATION TIME

| Algorithm Type | Maximum Errors or RMSEs [p.u.] | | CPU Time |
|---|---|---|---|
| | Real Part | Imaginary Part | |
| Proposed Method | $1.541\times10^{-3}$ | $1.451\times10^{-4}$ | 20.30 [ms] |
| WLS-based DSSE | $1.681\times10^{-3}$ | $1.842\times10^{-4}$ | 1000 trials: 174.79 [s] |

TABLE III
ESTIMATION PERFORMANCE IN ROBUSTNESS ANALYSIS

| Robustness Analysis | | Maximum errors at all nodes [p.u.] | | CPU Time [ms] |
|---|---|---|---|---|
| | | Real Part | Imaginary Part | |
| Voltage Profile | 0.90~0.95 | $1.652\times10^{-3}$ | $1.607\times10^{-4}$ | 20.43 |
| | 0.9~1.1 | $1.831\times10^{-3}$ | $1.793\times10^{-4}$ | 21.41 |
| Measurement Redun. | 1.221 | $1.330\times10^{-3}$ | $1.693\times10^{-4}$ | 18.30 |
| | 1.265 | $1.029\times10^{-3}$ | $1.642\times10^{-4}$ | 17.11 |

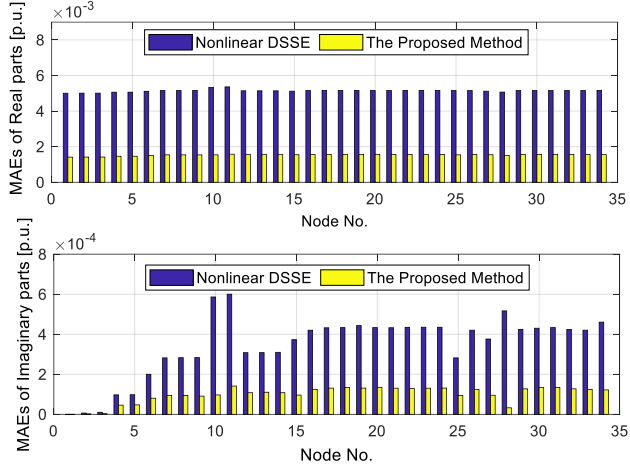

Fig. 2. Comparison in the estimation errors of the proposed method and the MAEs that may happen in a random sampling of measurements.

$$\boldsymbol{X}^{(j+1)} = (\boldsymbol{C}\boldsymbol{\mathcal{B}} + (\boldsymbol{I} - \boldsymbol{C}\boldsymbol{\mathcal{A}})\boldsymbol{X}^{(j)}) \cap \boldsymbol{X}^{(j)} \quad (17)$$

where $\varepsilon = 10^{-4}$. It is proved that starting from $\boldsymbol{X}^{(0)}$, the iterative process rapidly converges if $\|\boldsymbol{I} - \boldsymbol{C}\boldsymbol{\mathcal{A}}\| < 1$ according to the fixed point theorem [8], and $\|\cdot\|$ is any norm.

The final state estimate of $x$ in the proposed DSSE model (14) is obtained by $Mid[\boldsymbol{X}^{(j+1)}]$, considering that the measurement noises obey symmetric Gaussian distributions about zero means.

### III. NUMERICAL TEST

We test the proposed algorithm on the IEEE 34-bus distribution system [9]. The system is modified by adding four DGs at buses 822, 838, 856, and 864, respectively, and the installed capacity of each DG is 200 kVA. We obtain the true values of measurements and state variables by the power flow calculation, and all DG outputs are modeled as PQ buses with a constant power factor of 0.95 [2]. The maximum errors of measurements are set as 1% of these true values for the voltage magnitudes and powers from SCADA systems and 20% for pseudo-measurements. Table I displays the measurement placement scheme in the test system.

*A. Estimation Performance*

To evaluate the performance of the proposed method, we adopt the nonlinear WLS-based DSSE method in (1)-(5) as the baseline. In Fig. 2, the maximum absolute errors (MAEs) of the real and imaginary parts of voltages in 1000 MCSs are used to evaluate the estimation accuracy of the nonlinear WLS-based method in [2]. Also, using measurements from one of these MCSs, we calculate the errors of the estimated voltages in the proposed method. Fig. 2 implies that the WLS-based method in MCSs may produce the MAEs that reach up to around $5.34\times10^{-3}$ p.u., while the proposed method obtains accurate estimates with the maximum error $1.54\times10^{-3}$ p.u. The root mean square errors (RMSEs) of these estimated voltages in all MCSs are used to evaluate the overall estimation performance of this nonlinear DSSE method. Moreover, the maximums of the RMSEs are compared with the MAEs of the proposed method at all buses, shown in Table II. It concludes that in contrast to the WLS-based method in a Monte Carlo trial, the proposed method obtains the states more accurately. Also, the estimation accuracy of our method is close to that of the baseline method in 1000 MCSs, and however, the latter requires considerable sets of samplings and high computational cost.

Table II compares the computational efficiency of the proposed method with that of the WLS-based method. The CPU time of the proposed method accounts for about 12% of the average one that this nonlinear method takes in a single Monte Carlo trial, *i.e.*, 20.30 vs 174.79 milliseconds.

*B. Robustness Analysis*

This section discusses influence factors, such as operating condition and measurement redundancy, to illustrate the robustness of the proposed method.

1) Operating Condition

Considering the impacts of DG penetration on voltage profile, we investigate the estimation results of the proposed method in various operation ranges, *i.e.*, 0.90~0.95 p.u. and 0.9~1.1 p.u., shown in Table III. The comparison between these cases illustrates that a narrower voltage range around 1.0 leads to higher estimation accuracy and computational efficiency of the proposed method. Also, Table III validates the robustness of this method for various operating conditions.

2) Measurement Redundancy

The measurement redundancy shown in Table I is 1.176, and we further test the proposed algorithm with other measurement redundancies by adjusting the number of measurements and their locations. Table III gives the estimation performance of these tests, which shows that the efficacy of this method does not depend on the measurement arrangements. Also, the higher measurement redundancy leads to the overall improvement in estimation accuracy and computational efficiency.

## IV. CONCLUSION

This paper presents a highly efficient DSSE algorithm using the Taylor series of complex numbers and interval arithmetic techniques. Compared with the prior work [6] on interval arithmetic, which requires installing PMUs in distribution systems, this method provides a highly efficient substitute of the conventional nonlinear WLS-based methods without the use of PMU data. Numerical simulations illustrate the accuracy and efficiency of the proposed method in tackling nonlinear measurements with Gaussian noises.